# Non-Perturbative versus Perturbative Renormalization of Lattice Operators*

M. Göckeler[a,b], R. Horsley[c], E.-M. Ilgenfritz[c], H. Oelrich[d,a] H. Perlt[e], P. Rakow[f], G. Schierholz[d,a] and A. Schiller[e]

[a]Höchstleistungsrechenzentrum HLRZ, c/o Forschungszentrum Jülich, D-52425 Jülich, Germany

[b]Institut für Theoretische Physik, RWTH Aachen, D-52056 Aachen, Germany

[c]Institut für Physik, Humboldt-Universität, D-10115 Berlin, Germany

[d]Deutsches Elektronen-Synchrotron DESY, D-22603 Hamburg, Germany

[e]Institut für Theoretische Physik, Universität Leipzig, D-04109 Leipzig, Germany

[f]Institut für Theoretische Physik, Freie Universität Berlin, D-14195 Berlin, Germany

Our objective is to compute the moments of the deep-inelastic structure functions of the nucleon on the lattice. A major source of uncertainty is the renormalization of the lattice operators that enter the calculation. In this talk we compare the renormalization constants of the most relevant twist-two bilinear quark operators which we have computed non-perturbatively and perturbatively to one loop order. Furthermore, we discuss the use of tadpole improved perturbation theory.

## 1. Introduction

In [1] we have initiated a lattice calculation of the moments of the deep-inelastic structure functions of the nucleon, both for unpolarized and polarized beams and targets. We will not repeat the results nor the details of the calculation here, but we refer the reader to this work.

The program amounts to computing forward nucleon matrix elements of composite quark and gluon operators of definite twist and spin in a systematic fashion. Among the leading operators are:

$$\frac{1}{2}\sum_{\vec{s}} \langle \vec{p},\vec{s}|\mathcal{O}_{\{\mu_1\cdots\mu_n\}}(\mu)|\vec{p},\vec{s}\rangle$$
$$= 2v_n[p_{\mu_1}\cdots p_{\mu_n} - \text{traces}], \quad (1)$$

where

$$\mathcal{O}_{\mu_1\cdots\mu_n}$$
$$= \left(\frac{i}{2}\right)^{n-1} \bar{\psi}\gamma_{\mu_1}\overleftrightarrow{D}_{\mu_2}\cdots\overleftrightarrow{D}_{\mu_n}\psi - \text{traces}, \quad (2)$$

*Talk presented by G. Schierholz at *International Symposium on Lattice Field Theory*, Melbourne, July 1995

and

$$\langle \vec{p},\vec{s}|\mathcal{O}^5_{\{\sigma\mu_1\cdots\mu_n\}}(\mu)|\vec{p},\vec{s}\rangle$$
$$= \frac{1}{n+1}a_n[s_\sigma p_{\mu_1}\cdots p_{\mu_n} + \cdots - \text{traces}], \quad (3)$$

where

$$\mathcal{O}^5_{\sigma\mu_1\cdots\mu_n}$$
$$= \left(\frac{i}{2}\right)^n \bar{\psi}\gamma_\sigma\gamma_5\overleftrightarrow{D}_{\mu_1}\cdots\overleftrightarrow{D}_{\mu_n}\psi - \text{traces}. \quad (4)$$

Here $\{\cdots\}$ means symmetrization of the indices. The argument $\mu$ indicates that the operators are renormalized at the scale $\mu$.

The calculation is done in two steps. First one constructs bare lattice operators $\mathcal{O}(a)$, which belong to an irreducible representation of the hypercubic group $H(4)$ and in the limit of zero lattice spacing coincide with the classical euclidean continuum operators, and computes their matrix elements between nucleon states. The lattice operators are obtained from (2) and (4) by replacing the covariant derivative by the lattice covariant derivative (modulo factors of i)

$$\overrightarrow{D}_\mu(x,y) = \frac{1}{2}[U_\mu(x)\delta_{y,x+\hat{\mu}} - U_\mu^\dagger(x-\hat{\mu})\delta_{y,x-\hat{\mu}}]. \quad (5)$$

These operators are in general divergent as the lattice spacing goes to zero. In the second step the bare lattice operators are renormalized by defining

$$\langle q(p)|\mathcal{O}(\mu)|q(p)\rangle = \langle q(p)|\mathcal{O}(a)|q(p)\rangle\big|_{p^2=\mu^2}^{\text{tree}} \quad (6)$$

and

$$\langle q(p)|\mathcal{O}(\mu)|q(p)\rangle = Z_\mathcal{O}(\mu^2)\langle q(p)|\mathcal{O}(a)|q(p)\rangle, \quad (7)$$

where $|q(p)\rangle$ is a quark state of momentum $p$. In the continuum limit this prescription amounts to the momentum subtraction scheme. In the following we shall define the continuum quark fields to be $\sqrt{2\kappa}$ times the lattice quark fields.

In [1] we have seen that the nucleon matrix elements of the bare lattice operators can be computed relatively precisely. The renormalization constants $Z_\mathcal{O}$, on the other hand, were calculated in lattice perturbation theory to one loop order. It has been argued that the perturbative series converges slowly due to the contribution of tadpole diagrams [2]. In view of this we have computed the renormalization constants non-perturbatively [3] following a recent suggestion of [4]. In this talk we shall present some first results of our calculation and compare non-perturbative and perturbative values.

## 2. Non-Perturbative Renormalization

We shall restrict ourselves to the operators listed in Table 1. We have nothing to add to the perturbative calculation. A most complete list of perturbative renormalization constants is given in [1]. For earlier results and recent independent work see also [5].

The idea of the non-perturbative calculation is to compute the matrix element $\langle q(p)|\mathcal{O}(a)|q(p)\rangle$ in (7) directly on the lattice by numerical simulations. The only unpleasant feature in this calculation is that one has to fix the gauge. We have chosen the Landau gauge. This gauge suffers from the Gribov ambiguity. We sum over all Gribov copies.

| Operators | $\langle \mathcal{O} \rangle$ |
|---|---|
| $\mathcal{O}_{\{44\}} - \frac{1}{3}(\mathcal{O}_{\{11\}} + \mathcal{O}_{\{22\}} + \mathcal{O}_{\{33\}})$ | $v_2$ |
| $\mathcal{O}_{\{114\}} - \frac{1}{2}(\mathcal{O}_{\{224\}} + \mathcal{O}_{\{334\}})$ | $v_3$ |
| $\mathcal{O}_{\{1144\}} + \mathcal{O}_{\{2233\}} - \mathcal{O}_{\{1133\}} - \mathcal{O}_{\{2244\}}$ | $v_4$ |
| $\mathcal{O}_2^5$ | $a_0$ |
| $\mathcal{O}_{\{214\}}^5$ | $a_2$ |

Table 1

We consider the lattice bare Green function

$$G_\mathcal{O}(p) = \sum_{x,y} e^{-i(px-py)}\langle \psi(x)\mathcal{O}\bar{\psi}(y)\rangle. \quad (8)$$

The operators can be written $\mathcal{O} = \bar{\psi}\mathcal{V}\psi$. We define the free vertex operator $\mathcal{V}^{\text{tree}}$ by replacing the lattice covariant derivative by the ordinary derivative, which amounts to setting $U_\mu = 1$ in (5). The renormalization constant is then obtained by

$$Z_\mathcal{O}(\mu^2) = \frac{\text{Tr}\,\Gamma S(p)\mathcal{V}^{\text{tree}}(p)S(p)}{\text{Tr}\,\Gamma G_\mathcal{O}(p)}\bigg|_{p^2=\mu^2} Z_\psi(\mu^2) \quad (9)$$

for a suitably chosen projection matrix $\Gamma$. Here $S(p)$ is the quark propagator and $\sqrt{Z_\psi}$ is the renormalization constant of the fermion field.

The calculations are done on the same lattices as in [1]: $16^3 \cdot 32$ at $\beta = 6.0$ and for quenched Wilson fermions with $\kappa = 0.155, 0.153$ and $0.1515$. The results, using 125 configurations, are shown in Fig. 1 for $\kappa = 0.153$. We have labelled the $Z$'s by the operator matrix element they renormalize, and lvc stands for local vector current. If not stated otherwise, we set $a = 1$.

The momenta $p$ were chosen such that the individual components, $p_\mu$, differed by at most one unit of $2\pi/16$, so that even for our largest momenta none of the $p_\mu$ exceeded $\pi/2$. This should keep $O(a)$ effects small. On the other hand, for the method to work, the loop momenta squared should be much larger than the square of a typical hadron mass. Taking the glueball mass as a scale, this means that at $\beta = 6.0$ we can only expect to find agreement with perturbation theory for $2 \ll p^2$. The operators which involve higher powers of covariant derivatives are furthermore expected to be particularly sensitive to finite volume effects in the region of lower momenta.

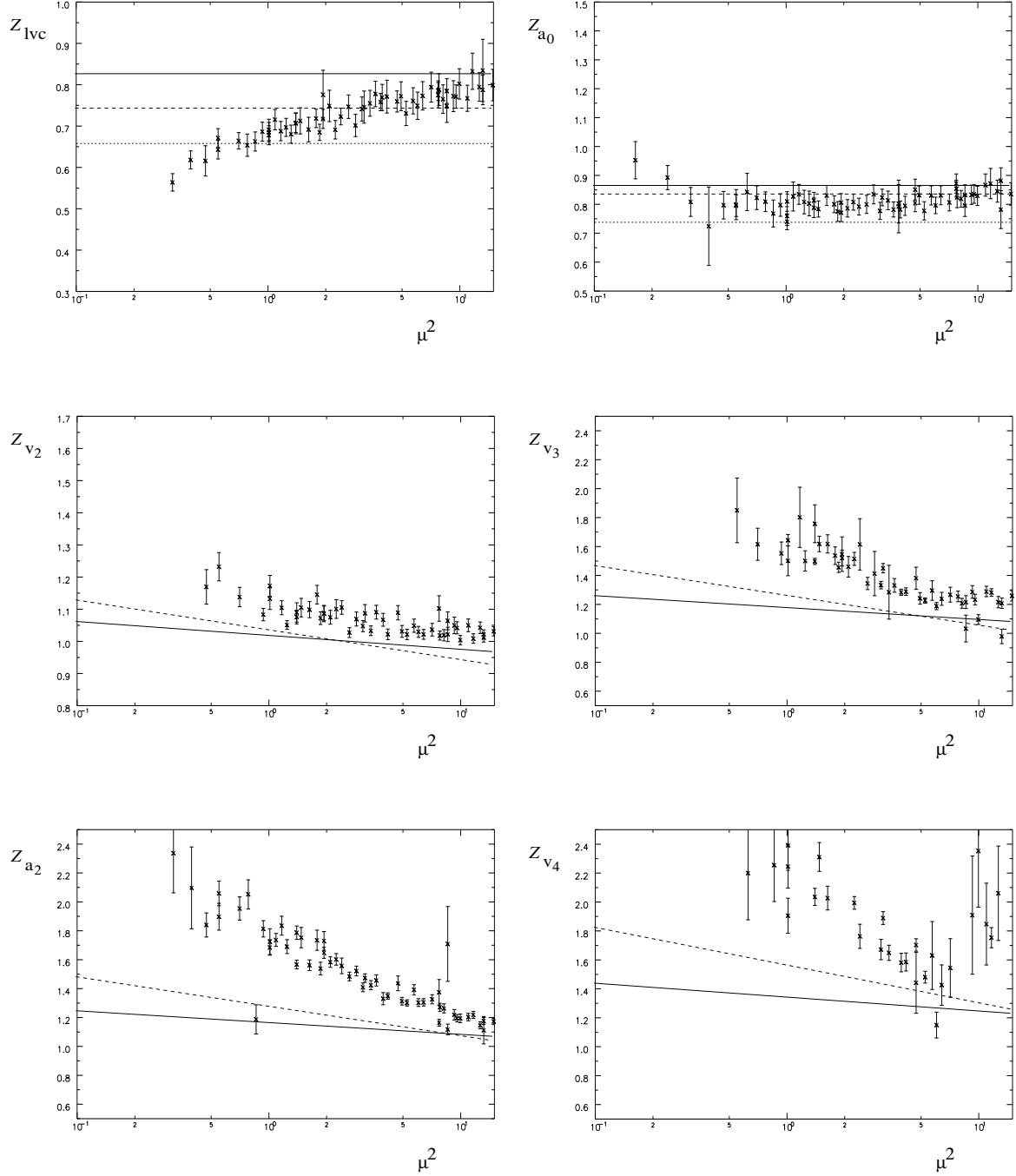

Figure 1. The renormalization constants $Z_{\mathrm{lvc}}$, $Z_{\mathrm{a}_0}$, $Z_{\mathrm{v}_2}$, $Z_{\mathrm{v}_3}$, $Z_{\mathrm{a}_2}$ and $Z_{\mathrm{v}_4}$ as a function of $\mu^2$ for $\kappa = 0.1530$. The solid lines are the perturbative results. The dotted and dashed lines are predictions of tadpole improved perturbation theory.

## 3. Comparison with Lattice Perturbation Theory

Let us now compare our results with the predictions of perturbation theory. The solid lines in Fig. 1 are the results of standard one-loop lattice perturbation theory [1]. There is considerable freedom in improving lattice perturbation theory. The dashed lines correspond to tadpole improvement [2], where we have taken $\tilde{\kappa}_c = \kappa_c u_0$ with $\kappa_c = 0.15693$ and $u_0 = 0.891$, while the dotted lines represent the choice $\tilde{\kappa}_c = 1/8$. In both cases the expansion parameter was taken to be $\alpha = 0.198$ [2]. The latter option (i.e. the dotted line) does not fit the data any better than standard perturbation theory. From $Z_{V_2}$ onwards it performs even worse, so that we shall not pursue this choice of parameters any further. In the case of $Z_{V_3}$ it has been found that the operator given in Table 1 mixes with the operator in another representation. The effect of mixing turned out to be very small though [1], so that we may safely neglect it here. We should also mention that for our choice of projection matrix $\Gamma$ the definition (9) does not precisely correspond to the renormalization prescription employed in perturbation theory. The two definitions can be converted into each other. The difference is insignificant.

The figures are organized according to the power of covariant derivatives in the operators. In one-loop perturbation theory $Z_{lvc}$ and $Z_{a_0}$ receive their major contribution from the leg tadpole diagrams, in $Z_{V_2}$ leg and operator tadpole diagrams cancel, in $Z_{V_3}$ and $Z_{a_2}$ the net contribution is (approximately) one power of $u_0$, and in $Z_{V_4}$ two powers. For $Z_{lvc}, Z_{a_0}$, and $Z_{V_2}$ the agreement with improved (the dashed line) and even standard perturbation theory is surprisingly close. However, from $Z_{V_2}$ onwards we find agreement only for larger momenta. It is also striking that the data show a somewhat steeper slope than the perturbative lines. (The slope originates from the anomalous dimensions of the operators.) This suggests perhaps that a more effective expansion parameter would be a value of $\alpha$ which is twice as large as the value used in the figures.

For the other $\kappa$ values, 0.155 and 0.1515, we found the same results, momentum for momentum, within the error bars. One might have expected the numbers to vary by a factor of [2,6] $1 + (1/\kappa - 1/\kappa_c)/2u_0$. This would have made a difference of nearly 10% between our largest and smallest $\kappa$ values. Our data do not support such a quark mass dependence.

## 4. Conclusions

We have seen that perturbative and non-perturbative renormalization constants agree within 20% for $\mu^2 \geq 4$ and better. At the present value of $\beta$ we expected the non-perturbative method to work for $\mu^2 \gg 2$.

Our data do not support taking $\tilde{\kappa}_c = 1/8$. We also do not see any dependence of the renormalization constants on the quark mass in the range of $\kappa$ values we have explored.

As far as we can tell, the agreement of perturbative and non-perturbative renormalization constants is no worse than in the case of improved fermions [4]. Wilson fermions, however, have the advantage that they suffer less from statistical fluctuations.